\documentclass[11pt,a4paper]{article}
\usepackage{jheppub}

\usepackage{pdflscape}
\usepackage{amsmath}
\usepackage{amssymb}
\usepackage{dcolumn}
\usepackage{bm}
\usepackage{color}
\usepackage{epsfig}
\usepackage{amsfonts}
\usepackage{graphicx}
\usepackage{subfigure}
\usepackage{dcolumn}

\newcommand{\be}{\begin{equation}}
\newcommand{\ee}{\end{equation}}
\newcommand{\bea}{\begin{eqnarray}}
\newcommand{\eea}{\end{eqnarray}}

\def\be{\begin{equation}}
\def\ee{\end{equation}}
\def\bea{\begin{eqnarray}}
\def\eea{\end{eqnarray}}

\begin{document}

\title{ Holographic Dark Energy from Fluid/Gravity
Duality Constraint by Cosmological Observations}

\author[a]{Behnam Pourhassan}
\author[b]{Alexander Bonilla}
\author[c]{Mir Faizal}
\author[b,d]{Everton M. C. Abreu}

\affiliation[a] {School of Physics, Damghan University, Damghan, 3671641167, Iran}
\affiliation[b] {Departamento de F\'isica, Universidade Federal de Juiz de Fora, 36036-330, Juiz de Fora, MG, Brazil}
\affiliation[c] {Irving K. Barber School of Arts and Sciences,  University of British Columbia
- Okanagan, Kelowna, BC V1V 1V7, Canada}
\affiliation[c] {Department of Physics and Astronomy, University of Lethbridge,
Lethbridge,   Alberta, T1K 3M4, Canada  }
\affiliation[d] {Grupo de F\'isica Te\'orica e Matem\'atica F\'isica, Departamento de F\'isica, Universidade Federal Rural do Rio de Janeiro,
23890-971, Serop\'edica, RJ, Brazil}

\emailAdd{b.pourhassan@du.ac.ir}
\emailAdd{abonilla@fisica.ufjf.br}
\emailAdd{f2mir@uwaterloo.ca}
\emailAdd{evertonabreu@ufrrj.br}

\abstract{In this paper, we obtain a holographic model of dark energy using the
fluid/gravity duality. This model will be dual to a higher
dimensional Schwarzschild black hole, and  we would use fluid/gravity duality
to relate to the parameters of this black hole  to such a cosmological model.
We will also analyze the   thermodynamics of such a solution, and  discuss the  stability model. Finally,
we  use cosmological data to constraint the parametric space of this
dark energy model.
Thus, we will use observational data  to  perform cosmography for this holographic model based
on fluid/gravity duality.}


\maketitle

\section{Introduction}
The holographic principle states that the number of degrees of freedom in a region of space area
equal to the degrees of freedom on the boundary of that region of space \cite{holo1, holo2}.
The AdS/CFT correspondence is a concrete realization of the holographic principle, as
it states that the string theory/supergravity on AdS spacetime
is dual to the superconformal field theory
on the boundary of that of that AdS spacetime
\cite{1, 2, 3}. An interesting part of the AdS/CFT correspondence is
that it can establish a duality between
weakly coupled theories and  strongly coupled coupled theories. Thus, it has been used to
study different aspects of strongly coupled theories which describe the
quark-gluon plasma (QGP) \cite{6,7,JHEP}, and this duality between the QCD and AdS is called
 AdS/QCD correspondence \cite{8,9}.  This correspondence has been used to study the field theory
 dual to STU model as a theory which could describe such physical systems \cite{11, 12, 13}.
 In fact, various   properties of QGP have been
studied using this duality with  STU black hole \cite{14,15, 16}.
The  AdS/CFT correspondence has also been used to study   condense matter systems,
and this holographic description of the condensed matter systems is called the
AdS/CMT correspondence \cite{17, 18, 19}. In fact, certain  deformations of the
AdS backgrounds have been used to holographically analyze  superfluid
\cite{20} and superconductor \cite{21}.

An interesting use of the AdS/CFT correspondence is that it can be used to holographically
analyze the   hydrodynamic description of strongly
coupled conformal field theories  \cite{22}.
This holographic description of the  hydrodynamic description of strongly
coupled conformal field theories is usually called the gauge/fluid duality.
The fluid/gravity duality is important, as it is hoped that certain problems in the
fluid mechanics such as the
 global regular solutions of the Navier-Stokes
equations and  turbulence phenomena could be analyzed holographically using this duality.
This duality can also be used to analyze fluids dual to certain black hole solutions.
In fact, it has been proposed  that
a five dimensional  Schwarzschild black hole is dual to an interesting fluid mechanical system
 \cite{23}. It has also been proposed that cosmological solution  can be studied holographically
 using this  duality \cite{24}.

The dark energy of the
 universe has also been  studied holographically \cite{h1, h2, h3, No1, h4}.
 Holographic dark energy with massive neutrinos has also been studied  and
  constraints using cosmological
 data from the Planck CMB lensing data, Planck CMB temperature data,
 the JLA supernova data, the baryon acoustic oscillation data,
 the cosmic shear data of weak lensing, the Hubble constant direct measurement,  and the
 redshift space distortions data \cite{h5}. The
 interacting holographic dark energy models  have been studied for
various different   interactions \cite{h6, h7}. It was  observed that
the type of interaction  terms is constraint by the cosmological data
 for these
interacting holographic dark energy models. Recently, it is demonstrated that any covariant gravity maybe described via
such holographic dark energy \cite{No2}. Thus, it is important and interesting to
study   models of dark energy based on the holographic principle.

We would like to clarify the various ways in which holographic principle can be used to study such a system.
The holographic principle can be used directly to the physical universe, as is done in most
models of holographic dark energy \cite{h1, h2, h3, No1, h4}.
It is also possible to consider systems which have the same degrees of freedom
as an AdS spacetime, and then use the holographic description to study such field theoretical systems.
This is the approach that has been used in condensed matter systems, where a specific condensed matter system
is analyzed using holography because the degrees of freedom describing such a system are the same as the degrees of
freedom of an AdS spacetime \cite{cond1, cond2}.
This is the approach we will use to analyze holographic dark energy
using fluid/gravity duality.  Thus, as dark energy can be analyzed as a fluid dynamical system
\cite{flud1, flud2, flud4, flud5},
and a fluid dynamical system can be analyzed using its gravity dual \cite{26, 25}, we will analyze the dark energy
using its gravity dual. We would like to point out that the main motivation for this work is that
just like condensed matter systems can be studied by analyzing their gravity duals \cite{cond1, cond2},
the dark energy can also be analyzed as a fluid mechanical system using its gravity dual.
 This is different from earlier works on holographic dark energy, where the holographic principle was directly applied
 to the physical universe.

 So, in this paper, we first note that fluid/gravity duality can be used to analyze various different
 fluid mechanical systems,
and this can be done by mapping the properties of those fluids
to an AdS spacetime    \cite{26, 25}. Then we observe that
it is possible to model the dark energy using fluid mechanical systems  \cite{flud1, flud2, flud4, flud5}.
So, we use the fluid/gravity duality   to map such a fluid mechanical systems, which can describe
dark energy, to an AdS solution.
We observe that the  gravity dual to such a fluid mechanical system is a
higher dimensional AdS-Schwarzschild black hole.
  We would like to point out that the AdS-Schwarzschild black hole is only used to obtain the dynamics of the fluid
mechanical system which is dual to it. However, after we obtain such a boundary description of a fluid,
we use cosmography to fix the values of  parameters in this system, so that it describes dark energy in our universe.
Thus, in this paper, we will use fluid/gravity duality to study the fluid dynamical properties of dark energy. It
would be interesting to extend this work further and analyze other  cosmological phenomena using such an approach.
This can be done by first  analyzing  such a  cosmological phenomena  using
a fluid dynamical system, and then finding a suitable gravity dual to such a fluid dynamical system.

So,  we use results fluid/gravity duality \cite{25} to analyze a  fluid dynamical model of  dark energy using
its gravitational dual.   This paper is organized as follows.
In section 2, we review procedure of obtaining equation of state of a fluid  using its gravity dual.
  In section 3, we use  the equation of state obtained in section 2, to construct dark energy model.
In section 4, we study thermodynamics of this dark energy model. In section 5, we use observational data to
fix the parameters in this model. Finally in section 6, we summarize our results in the conclusion.
So, in this paper, we start from fluid/gravity duality, and construct a model of dark energy using holography dual
to a AdS Schwarzschild black hole.
Then we study the thermodynamics of this holographic dark energy model, and finally constraint the parameters in this
model using observational data.

\section{Fluid/Gravity Duality}
In this section we will review the  fluid/gravity duality.
We first noted that at long-wavelengths,
the  effective dynamics of a continuum system can be described using
fluid  mechanics. Furthermore, according to fluid/gravity duality, this fluid mechanical system
on the boundary of an AdS spacetime  is dual to
 the bulk Einstein equations in the  AdS spacetime    \cite{26}.
 So, such a fluid mechanical system can be described  by an asymptotically
AdS spacetime  given by the following line element,
\begin{equation}\label{2}
ds^{2}=\frac{L^{2}}{r^{2}}\left(-f(r)dt^{2}+\frac{dr^{2}}{f(r)}+dX^{2})\right),
\end{equation}
where $L$ is the constant AdS radius, and
\begin{equation}
dX^{2}=dx_{1}^{2}+dx_{2}^{2}+dx_{3}^{2},
\end{equation}
is the three dimensional metric of the flat space. If we consider the five dimensional AdS Schwarzschild black hole with the event horizon radius $r_{h}$, then,
\begin{equation}\label{3}
f(r)=1-\frac{r^{4}}{r_{h}^{4}}.
\end{equation}
The action governed our model given as,
\begin{equation}\label{1}
S=\frac{1}{2\kappa}\int{d^{5}x\sqrt{-g}(R-2\Lambda)}+S_{Q}+S_{M},
\end{equation}
where $\kappa=8\pi G$, while  $S_{Q}$ and $S_{M}$
are boundary actions corresponding to Neumann and Dirichlet boundary conditions respectively.
It is indeed the FRW universe embedded in the five dimensional AdS Schwarzschild spacetime,
and the properties of the  fluid  can be holographically obtained  from the bulk  \cite{0102042}.
We will obtain an induced metric,  which resembles  the  FRW metric on a brane.
Now we  consider $u=\frac{1}{r}$, and  change the  variable,  so that the
coordinates $u$. Then, one can choose time parameter as $\tau$,  with the following condition,
\begin{equation}\label{emb1}
\frac{1}{h(u)}\left(\frac{du}{d\tau}\right)^{2}-h(u)\left(\frac{dt}{d\tau}\right)^{2}=-1,
\end{equation}
where $h(u)$ is given by
\begin{equation}\label{emb2}
h(u)=L^{2}u^{2}f(u),
\end{equation}
and $f(u)$ is given by
\begin{equation}\label{emb3}
f(u)=1-\frac{u_{h}^{4}}{{u}^{4}}.
\end{equation}
So,  the induced metric  takes the form of a  standard FRW metric,
\begin{equation}\label{emb4}
ds_{4}^{2}=-d\tau^{2}+u^{2}(\tau)d\Omega_{3}^{2}.
\end{equation}
It should be noted that the size of the four dimensional universe is specified by the radial distance, $u$, from the black hole center.

We can now analyze kinematic properties of the fluid using this fluid/gravity duality.
So, we can use the fluid/gravity duality \cite{drag1, drag2} to
obtain the equation of motion for this system
\begin{eqnarray}\label{EoM}
0&=&\frac{\partial}{\partial r}\left(\frac{f(r)}{r^{2}}\frac{y^{\prime}(r)}{\sqrt{L^{4}(1-\frac{\dot{y}(r)^{2}}{r^{2}f(r)}+\frac{f(r)}{r^{2}}y^{\prime}(r)^{2})}}\right)\nonumber\\
&+&\frac{1}{r^{4}f(r)}\frac{\partial}{\partial t}\left(\frac{\dot{y}(r)}{\sqrt{L^{4}(1-\frac{\dot{y}(r)^{2}}{r^{2}f(r)}+\frac{f(r)}{r^{2}}y^{\prime}(r)^{2})}}\right).
\end{eqnarray}
As from the time independent profile, we have  $\dot{y}(r)=0$, so equation of motion can be expressed as
\begin{equation}\label{EoM2}
\frac{\partial}{\partial r}\left(\frac{f(r)}{L^{2}r^{2}}\frac{y^{\prime}(r)}{\sqrt{1+\frac{f(r)}{r^{2}}y^{\prime}(r)^{2}}}\right)=0.
\end{equation}

The stress-energy tensor for this system can be written as \cite{Henningson, Balasubramanian},
\begin{equation}\label{4}
T_{ab}=-\frac{L^3}{\kappa r^{3}}\left(K_{ab}-Kh_{ab}+\Sigma h_{ab}-\kappa T_{ab}^{(R)}-\kappa T_{ab}^{(ct)}\right),
\end{equation}
where $h_{ab}$ is induced metrics on the hypersurface $Q$, and  $\Sigma$ is tension of $Q$. Here
$K$ is the trace of the extrinsic curvature
$K_{ab}$. In this system, $T_{ab}^{(R)}$ and $T_{ab}^{(ct)}$
are additional possible contributions from the intrinsic
curvature and counter-terms respectively. Here, it should be noted that stress-energy tensor (\ref{4}) is defined with respect to the intrinsic
hypersurface metric given by,
\begin{equation}\label{4-1}
\hat{h}_{ab}=\frac{r^{2}}{L^{2}}h_{ab}.
\end{equation}

\begin{figure}[h!]
 \begin{center}$
 \begin{array}{cccc}
\includegraphics[width=90 mm]{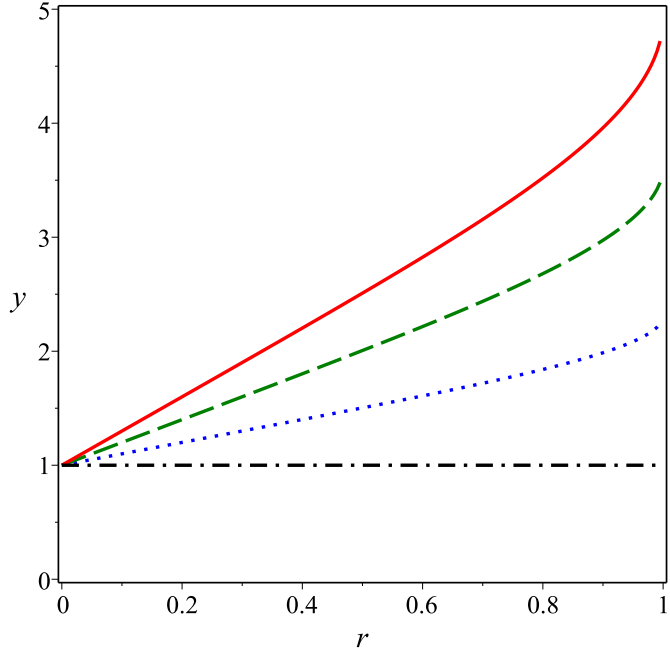}
 \end{array}$
 \end{center}
\caption{Hypersurface profiles in terms of $r$ for $y_{0}=1$ and
$r_{h}=1$. Here $c=0$ (red solid), $c=1$ (black dot), $c=2$ (black dash), $c=3$ (black dash dot).}
 \label{fig1}
\end{figure}

It should also be noted that following the Ref. \cite{25},  we consider the boundary specified by the condition of $x_{2}\equiv y = const$,
where $y$ is one of the coordinates on manifold $M$, and so we consider the AdS/BCFT \cite{Ali} on a half of Minkowski space.
It means that we consider problem in a half-space $y < 0$, hence one can parameterize a generic hypersurface $Q$ by the
profile $y(r)$. Then, for the simplicity, we consider only the first three terms in the Eq. (\ref{4}),  and use the following Neumann boundary
condition,
\begin{equation}\label{bound}
K_{ab}-(K-\Sigma)h_{ab}=8\pi GT_{ab}.
\end{equation}
Using the boundary stress-energy tensor,  the components of four dimensional boundary, can be written
as
\begin{eqnarray}\label{5}
p_{L}&=&\frac{L^{3}}{2\kappa r^{4}}\left(-2\Sigma L+\frac{y^{\prime}(r)(4f(r)-rf^{\prime}(r))}{(1+f(r)y^{\prime}(r)^{2})^{\frac{1}{2}}}\right),\nonumber\\
p_{T}&=&-\frac{\Sigma L^{4}}{\kappa r^{4}}\nonumber\\
&+&\frac{L^{3}}{2\kappa r^{4}}\left(\frac{2(2f(r)-rf^{\prime}(r))y^{\prime}(r)+f(r)(4f(r)-rf^{\prime}(r))^{\prime}y^{\prime}(r)^{3}-2rf(r)y^{\prime\prime}(r)}{(1+f(r)y^{\prime}(r)^{2})^{\frac{3}{2}}}\right),
\nonumber\\
\rho&=&\frac{L^{3}}{2\kappa r^{4}}\left(2\Sigma L+\frac{(rf^{\prime}(r)-4f(r))y^{\prime}(r)-4f(r)^{2}y^{\prime}(r)^{3}+2rf(r)y^{\prime\prime}(r)^{2}}{(1+f(r)y^{\prime}(r)^{2})^{\frac{3}{2}}}\right),
\end{eqnarray}
where $\rho$ is energy density. Here, $p_{T}$ and $p_{L}$ are transverse and
longitudinal pressures, respectively. The prime denotes a derivative with respect to $r$.
Neumann boundary conditions require $T_{ab}=0$ which yields to $\Sigma=0$ (for the dimensions greater that three).
In fact,  it is required to set  $\Sigma=0$
to obtain a perfect fluid equation of state.
It is also possible to   set $\Sigma L = 2 cos\theta$, where $\theta$ is the angle of
the hypersurface $Q$ with the $y$-axis. As the boundary for this spacetime   can be described by the generic hypersurface $Q$,
such that it is  parameterized by the function $y(r)$. In order to have
fluid mechanical  system, we require  isotropic pressures, and so we have
\begin{equation}\label{6}
p_{T}=p_{L}=p
\end{equation}
Hence, we can obtain
\begin{equation}\label{8}
\sqrt{f(r)}y^{\prime}(r)=c,
\end{equation}
where $c$ is an integration constant. Thus, we can write
\begin{equation}\label{9}
y-y_{0}=c\int_{0}^{r}{\frac{dr}{\sqrt{f(r)}}}.
\end{equation}
It is important relation from holographic point of view which is relation between radial coordinate in gravity side
and normal space coordinate in fluid side.
In the case of AdS Schwarzschild black hole we can solve integration (\ref{9}) numerically and plot
$y$ in terms of $r$ in the Fig. \ref{fig1}. Behavior of $y$ is completely agree with the results of the what is expected
form a  four dimensional Schwarzschild black hole  \cite{25}.
In fact, as we have to only consider the
  positive value of $c$,  so we have  $c\geq0$ \cite{25}.

In the case of $y(r_{h})=y_{h}$, we can write,
\begin{equation}\label{10}
\tan\theta\approx\frac{\Delta y}{r_{h}},
\end{equation}
where $\Delta y=y_{h}-y_{0}$ is characteristic distance scale, and $\theta$ is angle of the black curves with the red solid line.
Very close to boundary, where $r\ll r_{h}$ (close to boundary),  one can use the Taylor expansion to obtain,
\begin{equation}\label{11}
y-y_{0}\approx 2^{-\frac{3}{4}}c r_{h}\left[\tan^{-1}{\left(\frac{r}{2^{\frac{1}{4}}r_{h}}\right)}+\frac{1}{2}\ln{\left(\frac{r+2^{\frac{1}{4}}r_{h}}{r-2^{\frac{1}{4}}r_{h}}\right)}\right].
\end{equation}
Now,  we can write the pressure of this system as
\begin{equation}\label{13}
p=\frac{L^{3}}{2\kappa r^{4}}\left(-2\Sigma L-\frac{c(rf^{\prime}(r)-4f(r))}{\sqrt{(1+c^{2})f(r)}}\right).
\end{equation}
We can also write the energy density of this system as,
\begin{equation}\label{12}
\rho=\frac{L^{3}}{2\kappa r^{4}}\left(2\Sigma L
-\frac{4cf(r)(1+c^{2})}{\sqrt{(1+c^{2})f(r)}}\right).
\end{equation}
Thus, it is possible to mathematically  analyze this fluid mechanical system using its gravity dual.
As it is known that the dark energy can be modeled using a fluid mechanical system, and this fluid mechanical
system can be analyzed using its holographic dual, we will use
  use this holographic description of a fluid to describe  a   dark energy
model, and holographically analyze such a model. It may be noted  the holographic description would be used to mathematically
analyze the fluid mechanical aspects of this system using the fluid/gravity duality. Its connection to the
FRW universe will be made by assuming that this fluid describes a dark energy model in the Friedmann equation,
and it is a realistic model, as it is constrained by observational data.


\section{Holographic Dark Energy}
In this section, we will use to holographic description to analyze a holographic model of
dark energy dual to such a background geometry.
So, here we will model dark energy with a perfect fluid, with
  energy density $\rho$ and pressure $p$. The equation of state for this model of dark energy
  would be
\begin{equation}\label{14}
\omega=\frac{p}{\rho}.
\end{equation}
Now using Eqs. (\ref{6}), (\ref{12}) and (\ref{13}), we observe that for
 $\omega=-1$, we have  the  equation of state with a cosmological constant
\begin{equation}\label{15}
p=-\rho=\frac{L^{3}}{2\kappa r^{4}}\left(-2\Sigma L+4c\sqrt{(1+c^{2})f(r)}\right).
\end{equation}
It may be noted that for the general case,  we should set $\Sigma=0$.
  Thus,  comparing Eqs.  (\ref{12}) and (\ref{13}), and using  (\ref{14}), we obtain
\begin{equation}\label{16}
\frac{f^{\prime}(r)}{f(r)}=\frac{4}{r}[1+\omega(1+c^{2})].
\end{equation}
Solution of this equation can be expressed as
\begin{equation}\label{17}
f(r)=Ar^{N},
\end{equation}
where $A$ is an integration constant, and $N=4(1+(1+c^{2})\omega)$.\\
The Friedmann equation in flat space for this  dark energy model can be written as
\begin{equation}\label{18}
H^{2}=\frac{\kappa}{3}\rho,
\end{equation}
where $H=\frac{\dot{a}}{a}$ is the Hubble's expansion parameter and
  $a(t)$ is the scale factor. Also dot denotes derivative with respect to the cosmic time. In the case of matter dominant we can write
\begin{equation}\label{20}
a=a_{0}t^{\frac{2}{3(1+\omega)}},
\end{equation}
and so  $H=\frac{2}{3(1+\omega)t}$. If we set $A=-r_{h}^{-4}$, and
assume that  $r\gg r_{h}$, then  we obtain,
\begin{equation}\label{21}
r_{h}=\sqrt{\frac{3}{2}}(1+\omega)c_{1}L^{\frac{3}{2}}\frac{t}{r_{b}},
\end{equation}
where $c_{1}$ is an arbitrary constant. Now
$r_{b}$ is the radius at which the  fluid
describes the  dark energy, such that
\begin{equation}\label{22}
Ar^{4(1+(1+c^{2})\omega)}+r_{h}^{-4}r^{4}=1
\end{equation}
where $r_{b}$ is the real positive root of this equation.
It may be noted that we have now obtained a
time-dependent black hole with the horizon given.
Hence, we conclude that $r_{h}$ is constant and  $r_{b}$ is time-dependent. So,
$r_{b}$  corresponds to time-dependent dark energy. It means that certain sheet of fluid with fixed $y$ behaves as dark energy.
Hence, we rewrite the equation (\ref{21}) as
\begin{equation}\label{21-re}
r_{b}=C\frac{t}{r_{h}},
\end{equation}
where $C\equiv(\sqrt{\frac{3}{2}}(1+\omega)c_{1}L^{\frac{3}{2}})^{-1}$ is a constant.\\
It may be noted that if  we assume $1+(1+c^{2})\omega=2$, then it is possible to
 solve the equation (\ref{22}) and obtain,
\begin{equation}\label{23}
r_{b}^{4}=\frac{r_{h}^{-4}}{2A}(-1+\sqrt{1+4Ar_{h}^{8}}).
\end{equation}
In the case of $c=\sqrt{2}$, we  obtain $\omega=\frac{1}{3}$, and this  is the equation of state
of ultra-relativistic matter.  This can represent
radiation matter in the very early universe. It may be noted that for any value of positive $A$, one can obtain $r_{b}\leq r_{h}$. For the large values of  $r_{h}$, we have
$r_{b}\rightarrow1$, and we can   obtain
time-dependent $r_{b}$ as,
\begin{equation}\label{23-1-1}
r_{b}=\sqrt{\frac{Ct}{1+AC^{4}t^{4}}},
\end{equation}
where $C$ is constant.\\
Now if  $(1+c^{2})\omega=-\frac{1}{2}$, then we can solve the equation (\ref{22}) and obtain,
\begin{equation}\label{24}
r_{b}^{2}=\frac{Ar_{h}^{4}}{2}(-1+\sqrt{1+\frac{4r_{h}^{-4}}{A^{2}}}).
\end{equation}
The  expansion of the universe is accelerating for any equation of state
$\omega < - \frac{1}{3}$ . So,
 for any value of positive $A$,  we have $r_{b}\leq r_{h}$. Now we  can write the
time-dependent radius   as
\begin{equation}\label{24-0}
r_{b}=(1+\sqrt{1+64C^{4}t^{4}})^{\frac{1}{4}}.
\end{equation}
In the  large $t$ limit, we obtain  $r_{b}\propto \sqrt{t}$. However,
the energy density of this case is negative; hence it is un-physical.\\
Now for $(1+c^{2})\omega=-2$, we can  solve the Eq.  (\ref{22}) and obtain,
\begin{equation}\label{24-1}
r_{b}^{4}=\frac{r_{h}^{4}}{2}(1\pm\sqrt{1-4Ar_{h}^{-4}}).
\end{equation}
Here for  $0\leq c^{2}\leq1$,  we have $\omega\leq-1$,
and we get  the hypothetical phantom energy and would cause a big rip.
So,  for any value of positive $A$,  we have $r_{b}\leq r_{h}$.
Now from  Eq.  (\ref{24-1}), we obtain
\begin{equation}\label{24-1-2}
r_{b}=\frac{1}{6}[216X(t)+\frac{2592C^{4}t^{4}}{X(t)}]^{\frac{1}{4}},
\end{equation}
where
\begin{equation}\label{24-1-3}
X(t)=\left(-108AC^{4}t^{4}+12\sqrt{-12C^{12}t^{12}+81A^{2}C^{8}t^{8}}\right)^{\frac{1}{3}}.
\end{equation}
It is easy to find that $r_{b}$ becomes proportional to $\sqrt{t}$ at the large $t$, otherwise we get negative density.\\
In general, using the conditions  given by Eqs. (\ref{16}) and (\ref{22}),  we obtain ,
\begin{equation}\label{25}
p=\omega\rho=\frac{2c(1+c^{2})\omega L^{3}}{\kappa r_{b}^{4}}f(r_{b}).
\end{equation}
As discussed above, using explicit time-dependent $r_{b}$,
and Eq.  (\ref{20}), we  obtain
\begin{equation}\label{25-0}
\rho=C_{1}a^{-6(1+\omega)}+\rho(0),
\end{equation}
where

\begin{equation}
C_{1}\equiv \frac{2L^{3}c(1+c^{2})r_{h}^{4}}{\kappa a_{0}^{-6(1+\omega)}}
\end{equation}
is a constant and $\rho(0)$ is initial density at $a=0$. In that case the scale factor obtained as follow,
\begin{equation}
a=\left(\frac{C_{2}\rho(0)e^{4\rho(0)(1+\omega)t}-C_{1}}{\rho(0)}\right)^{\frac{1}{6(1+\omega)}},
\end{equation}
where $C_{2}$ is an integration constant.
In this section, we model dark energy using a holographic fluid. In the next section, we
will study thermodynamics of this holographic  fluid representing the
 dark energy. Finally it should be note that to have consistent dimensional analysis we set $4\pi G=1$ which means $\kappa=2$ is dimensionless parameter.


\section{Thermodynamics analysis}
Now, we can analyze the thermodynamics of this holographic dark fluid.
We first observe that
Eqs.  (\ref{2}) and (\ref{3}) describe a black hole with the horizon radius $r_h$. General relation for the black hole Hawking temperature given by,
\begin{equation}
T_{H}=\frac{1}{4\pi}f^{\prime}(f)|_{r=r_{h}},
\end{equation}
and it has relation with local temperature on the surface $Q$ as
\begin{equation}
T=\frac{T_{H}}{\sqrt{f(r)}}.
\end{equation}
So we can write  the Hawking temperature for this black hole as
\begin{equation}\label{26}
T_{H}=\frac{1}{\pi r_{h}}.
\end{equation}
Also the local temperature on the surface $Q$ given by the following expression,
\begin{equation}\label{27}
T=\frac{r_{h}}{\pi\sqrt{r_{h}^{4}-r^{4}}}.
\end{equation}
So,  the local temperature where dual fluid  describes the dark energy  is given by the following expression,
\begin{equation}\label{28}
T_{b}=\frac{r_{h}}{\pi\sqrt{r_{h}^{4}-r_{b}^{4}}}.
\end{equation}
In the case of near boundary ($r_{b}\rightarrow0$), the  local temperature is equal
to the Hawking temperature, $T_{b}=T_{H}$.
The local entropy density for this system can be written as
\begin{equation}\label{29}
s=\frac{p+\rho}{T}.
\end{equation}
So,  by using Eqs.,  (\ref{25}) and (\ref{28}), we   obtain the
entropy of the fluid corresponding to the dark energy,
\begin{equation}\label{30}
s=c_{\omega}\frac{(r_{h}^{4}-r_{b}^{4})^{\frac{3}{2}}}{r_{b}^{4}r_{h}^{5}},
\end{equation}
where
\begin{equation}
c_{\omega}=2\pi L^{3}c(1+c^{2})(1+\omega),
\end{equation}
is a constant.
To have positive entropy,  we should have $\omega>-1$, otherwise it may produce
phantom energy. It may be noted that
  with the current observational data, it is impossible
to distinguish between phantom $\omega < - 1$ and non-phantom $\omega \geq - 1$.
It is clear that the entropy density is constant on the surface,
which is proportional to the area of
the horizon swept by the hypersurface.\\
Now using the Eqs.  (\ref{12}), (\ref{13}), and (\ref{29}), we   obtain the following entropy of the horizon,
\begin{equation}\label{32}
s(r=r_{h})=\frac{2\pi L^{3}c}{\sqrt{1+c^{2}}}\frac{1}{r_{h}^{3}}.
\end{equation}
It is easy to check that for the $c\gg1$ the above equation produces the five dimensional Schwarzschild entropy.\\
Furthermore, using the Eq.  (\ref{30}), we obtain Helmholtz free energy as,
\begin{equation}\label{33}
F=-\int{sdT},
\end{equation}
while specific heat for this system can be written as
\begin{equation}\label{36}
C_{v}=T\frac{\partial s}{\partial T}.
\end{equation}
So we can write the  Helmholtz free energy of the surface as
\begin{equation}\label{34}
F=-\frac{c_{\omega}(r_{b}^{4}-4r_{h}^{4}\ln{r_{h}})}{4\pi r_{h}^{4}r_{b}^{4}}.
\end{equation}
while the specific heat at the surface,  $r=r_{b}$, given by
\begin{equation}
C_{v}=-\frac{5c_{\omega}(r_{b}^{4}+\frac{r_{h}^{4}}{5})(r_{h}^{4}-r_{b}^{4})^{\frac{3}{2}}}{r_{h}^{5}r_{b}^{4}(r_{b}^{4}+r_{h}^{4})}.
\end{equation}
This specific heat is negative for $\omega>-1$. Now, we can use the specific heat and speed of sound  to study the  stability of the model.
There is another test for stability of the model, and this test is based on the  speed of sound,
\begin{equation}\label{37}
v_{s}^{2}=\frac{\dot{p}}{\dot{\rho}}=\frac{d p}{d\rho}.
\end{equation}
For the fluid  describing holographic dark energy, we have
\begin{equation}\label{38}
v_{s}^{2}=\omega.
\end{equation}
So,  if $\omega>0$, we get $v_{s}^{2}>0$ and  $C_{v}<0$. However,
if $\omega<-1$, then we get  $C_{v}>0$ and  $v_{s}^{2}<0$.


\section{Cosmology}
In this section, we briefly describe the observational data sets, the fitting method
used to constrain the parameters of
our model based on Friedmann equation (\ref{18}) in the presence of an effective
dark energy in the form of (\ref{25-0}). A cosmography analysis, comparison with observational data like \cite{2014A&A...571A..16P, BarbozaJr201219} and
the evolution of model also are presented in what follows.

\subsection{Data Sets and Fitting Method}

In order to constrains the free parameters of the models, we use i) The Union $2.1$ sample  \cite{2012ApJ...746...85S},
which contains $580$ Supernovae type Ia (SNIa), and we fit the SNIa data by minimizing the $\chi^2$ value as defined in \cite{metodo}. ii) The baryon acoustic oscillations (BAO) measurements from WiggleZ BAO data,
where the total $\chi^2$ for all the BAO data sets have 3 data point ($\bar{A}_{obs} = (0.447, 0.442, 0.424)$ at $z=(0.44,0.60,0.73)$)
from \cite{2005ApJ...633..560E}. iii) We adopt also 36 Observational Hubble Data (Hz) at different redshifts ($0.0708 \leq z \leq 2.36$) obtained
from \cite{2015arXiv150702517M}, where 26 value are deduced from the differential age method, whereas 10 corresponds
to that obtained from the radial BAO method. The $\chi^2$ function for Hz is built as in \cite{metodo}.\\

In order to find the value parameters for a given statistic we need maximum likelihood $\mathcal{L}_{max}$ as a function of the best fit parameters $p_{i}^{m}$, which given by,
\begin{equation}\label{40}
 \mathcal{L}_{max}(p_{i}^{m})= exp \left[ -\frac{1}{2}{\chi_{min}^{2}(p_{i}^{m})} \right]
\end{equation}
We use Gaussian errors distribution of maximum likelihood where, $\chi_{min}^{2}(p_{i}^{m})=-2 \ln \mathcal{L}_{max}(p_{i}^{m})$, \cite{2010arXiv1012.3754A}. Hence we have, $\chi_{min}^{2}= \chi^{2}_{SNIa} +\chi^{2}_{BAO}+\chi^{2}_{H(z)}$. Such Gaussian uncertainties of parameters given by the following Fisher matrix
\begin{equation}\label{41}
F_{ij}=\frac{1}{2}\frac{\partial ^2 \chi_{min}^{2}}{\partial p_i \partial p_j},
\end{equation}
which will be use to constraint the cosmological parameters from observational
data \cite{2009arXiv0901.0721A,2012JCAP...09..009W}.  $p_i$ and $p_j$ in the equation (\ref{41}) are the free parameters of the given model.  The covariance matrix $\left[ C_{cov} \right]$  is the inverse of the Fisher matrix, and the uncertainties expressed by,

\begin{equation}\label{41-1}
\sigma_i = \sqrt{Diag \left[ C_{cov} \right]_{ij} }.
\end{equation}

\begin{figure}[h!]
\begin{center}$
\begin{array}{cccc}
\includegraphics[scale=1]{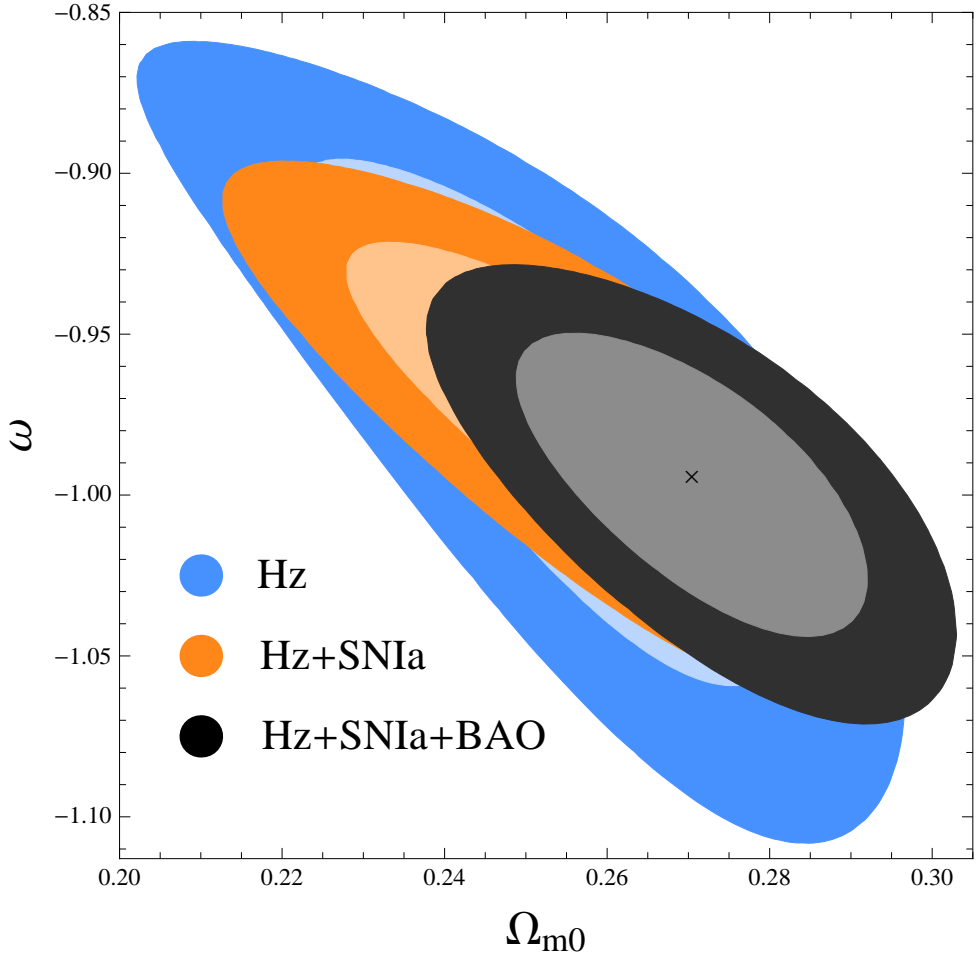}
\end{array}$
\end{center}
\caption{1$\sigma$ and 2$\sigma$ confidence regions of the Fluid/Gravity model obtained from the joint analysis $H(z)$, $H(z)$ $+$ SNIa and  $H(z)$ $+$ SNIa $+$ BAO for the free parameters ($h$, $\Omega_m$, $w$).}
\label{Model A}
\end{figure}

Figure (\ref{Model A}) shows the $68.27\%$ and $95.45\%$ regions of statistical confidence in the plane ($\Omega_{m0}-w$)
considering the observational data from $H(z)$, $H(z)+ SNIa$ and  $H(z) + SNIa + BAO$. Note that these results are in agreement with
$\Lambda$CDM model, which is obtained for $w=-1$ is inside the obtained region. On the other hand, the equation of state (EoS)
can help us to classify the model as phantom if EoS $w < -1$ , or as quintessence if $w > -1$.
Here, we note that the EoS is closed at $w = -1$ for all combinations of data. But, the introduction of $SNIa + BAO$ on $H(z)$
causes small variation into direction to a possible phantom dynamic.
In all combinations of data, although within the margin of error this cannot yet be still discriminated.
Table \ref{M_A} summarize the main results of the statistical analysis, and we can notice that although the value of $\chi^2_{min}$ for
$H(z)$ is a little below 36, this is greatly improved when the other data sets are introduced. Hence in summary, it is clear that Fluid/Gravity
model, under all the above three different combinations of statistical data sets, remains close to $\Lambda$CDM cosmology.

\begin{table*}[!h]
\begin{center}
\begin{tabular}{ccccc}
\hline
\hline
Data                   & $h$ & $\Omega_m$ & $w$ &  $\chi^2_{min}$ \\
\hline
$H(z)$                   & $0.698 \pm 0.042$ &  $0.254 \pm 0.021$ &  $-0.972 \pm 0.111$ & 17.151 \\

$H(z)$+SNIa          & $0.687 \pm 0.013$  & $0.256 \pm 0.017$   & $-0.973 \pm 0.029$ & 579.484 \\

$H(z)$+SNIa+BAO & $0.693 \pm 0.012$ &  $0.270 \pm 0.014$  & $-0.994 \pm 0.025$  & 581.293 \\
\hline
\hline
\end{tabular}
\end{center}
  \caption{Summary of the best fit values at $1\sigma$ of the free parameters ($h$, $\Omega_m$, $w$) to Fluid/Gravity model, for three different observational data sets with  $\chi^2_{min}$.}
      \label{M_A}
\end{table*}

\subsection{Cosmography}
In order to extract some observational parameters we use a Taylor series of the scale factor $a(t)$ around the current time ($t_0$) as follow \cite{2012PhyU...55A...2B},
\begin{equation}\label{41-1}
\frac{a(t)}{a(t_0)} = 1 + \frac{H_0}{1!} \left[ t - t_0 \right]  - \frac{q_0}{2!} H^2_0 \left[ t - t_0 \right]^2 +
\frac{j_0}{3!} H^3_0 \left[ t - t_0 \right]^{3}+\cdots,
\end{equation}
where
\begin{eqnarray}\label{42}
H(t) &\equiv& \frac{1}{a}\frac{da}{dt}; \nonumber\\
q(t) &\equiv& -\frac{1}{a}\frac{d^2 a}{dt^2} H(t)^{-2}; \nonumber\\
j(t) &\equiv& \frac{1}{a}\frac{d^3 a}{dt^3} H(t)^{-3},
\end{eqnarray}
The first term is Hubble expansion parameter $H(t)$, the second is called deceleration parameter $q(t)$ which could determine accelerating or decelerating of the universe, and the last term is called jerk parameter $j(t)$. The big advantage of this method is that
we can investigate the cosmic expansion without assuming any modification of gravity theory or dark energy model due to its geometrical
approach. The dependence between the free parameters of Fluid/Gravity model and the kinematic parameters $H(z)$, $q(z)$ and $j(z)$
can be obtained from the equations (\ref{18}) and  (\ref{25-0}) as

\begin{equation}\label{43}
H(z) = H_0  \left( \Omega_{r0} (1+z)^4 + \Omega_{m0}(1+z)^3 + \Omega_X (z) \right)^{1/2},
\end{equation}
where $\Omega_X (z)= (1-\Omega_{m0}-\Omega_{r0})(1+z)^{6(1+w)}$ and $\Omega_X (z)= \rho (z) /\rho_{cri}$ is given by equation (\ref{25-0}),
$\rho_{cri}= 3 H_0^2 / 8\pi G$ is critical density, $\Omega_{r0}$ and $\Omega_{m0}$ are radiation and matter density respectively,
and $w$ is the EoS. Here, we have that $\Omega_{r0}(h)=\Omega_{\gamma} (h) (1 + 0.2271 N_{eff})$,
where $\Omega_{\gamma} (h) = 2.469 10^{-5} h^{-2}$ is the density of photons, $N_{eff}=3.046$ is the effective number of neutrino species
\cite{2014A&A...571A..16P}, and $h=H_{0}/100 \mathrm{kms}^{-1}\mathrm{Mpc}^{-1}$ is dimensionless Hubble parameter.
Thus, our model has 3 parameters free $\left\lbrace h,\Omega_{m0},w\right\rbrace $, whose fit with the observational data
is shown in Table \ref{M_A}.

\begin{figure}[h!]
\begin{center}$
\begin{array}{cccc}
\includegraphics[scale=1]{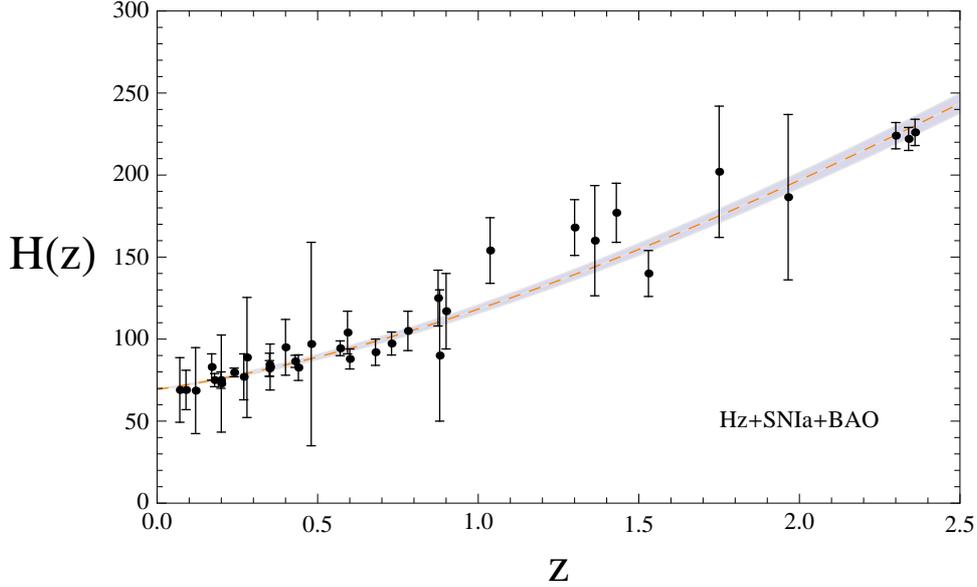}
\end{array}$
\end{center}
\caption{Using data set we plot the $H(z)$ function reconstructed using the best fit values for the $H(z) + SNIa + BAO$ case and the
Observational Hubble data set . We consider the error propagation at  $1\sigma$ (Gray region) in the best fit parameters ($h$, $\Omega_m$, $w$).}
\label{Model B}
\end{figure}

The deceleration and jerk parameters \cite{BarbozaJr201219} are obtained as

\begin{eqnarray}\label{44}
 q(z) &=& -1+\frac{(1+z)}{H(z)}\frac{dH(z)}{dz};\nonumber\\
 j(z) &=& q^2 + \frac{(1+z)^2}{H(z)}\frac{d^2H(z)}{dz^2}.
\end{eqnarray}

\begin{figure}[h!]
\begin{center}$
\begin{array}{cccc}
\includegraphics[scale=1]{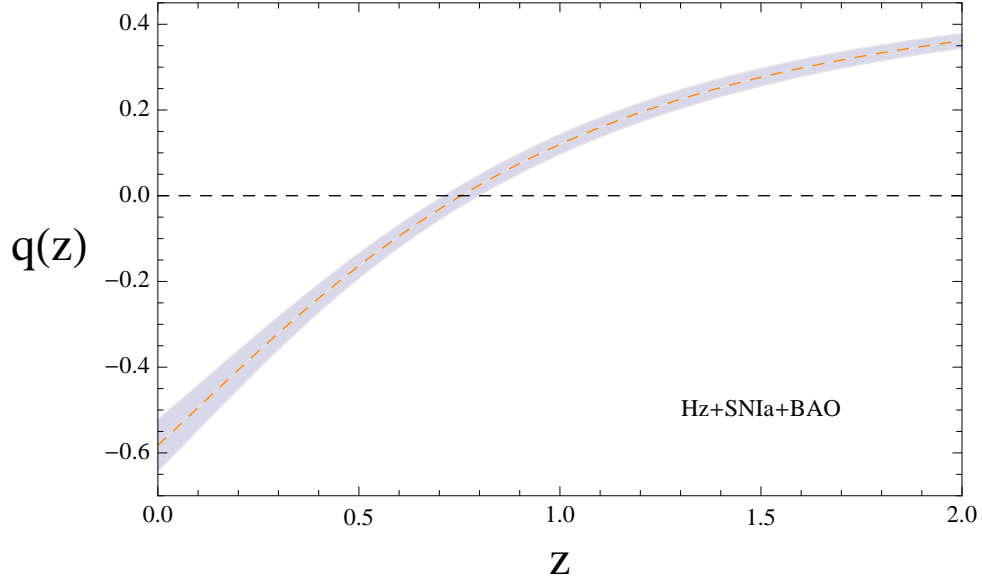}
\end{array}$
\end{center}
\caption{Using data set we plot the deceleration parameter reconstructed using the best fit values for the $H(z) + SNIa + BAO$ case. We consider the error propagation at $1\sigma$ (Gray region) in the best fit parameters ($h$, $\Omega_m$, $w$).}
\label{Model C}
\end{figure}

In the case of $q(z)>0$, and $\ddot{a}(z)<0$; the universe expansion decelerated, which is expected. In the case of sufficiently large  $\Omega_X$ ($\Omega_X >\Omega_m $), we have negative deceleration parameter and $\ddot{a}(z)>0$, which is corresponding to an accelerated expansion of the universe as illustrated by the Figure (\ref{Model C}) in agreement with current observational data.\\

\begin{figure}[h!]
\begin{center}$
\begin{array}{cccc}
\includegraphics[scale=1]{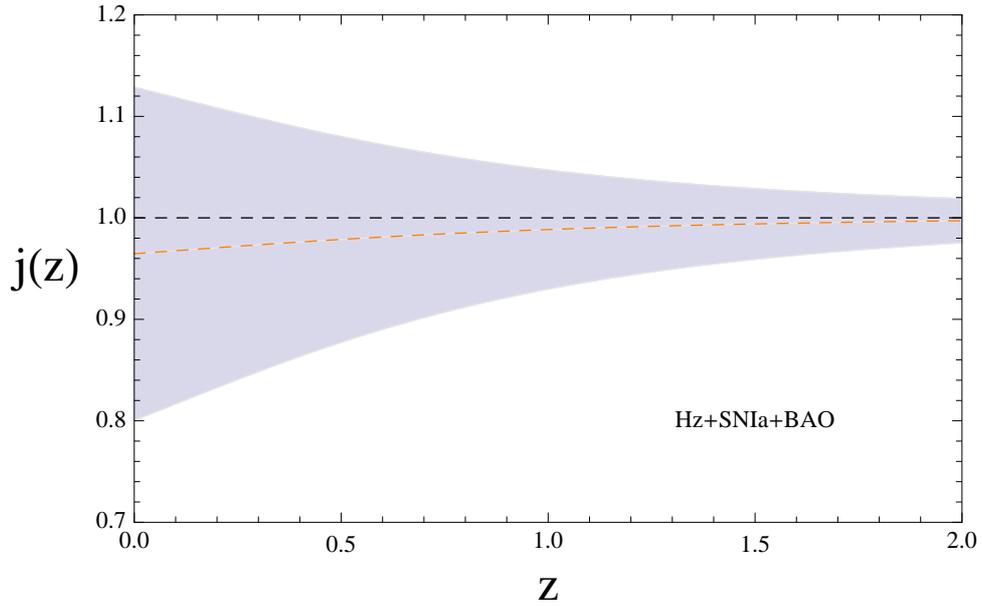}
\end{array}$
\end{center}
\caption{Using data set we plot the $j(z)$ parameter reconstructed using the best fit values for the $H(z) + SNIa + BAO$ case. We consider the error propagation at $1\sigma$ (Gray region) in the best fit parameters ($h$, $\Omega_m$, $w$).}
\label{Model D}
\end{figure}

Figure (\ref{Model B}) shows the evolution of $H(z)$ obtained in our analysis with propagation of error to $1\sigma$ (gray region)
obtained from the best fit of parameters with all observational data, which seems agree with this data set. Figure (\ref{Model C}) shows
the deceleration parameter $q(z)$ using all data set and how is expected our model show $q(z) < 0$ at late times and $q(z) > 0$
at earlier epoch. It means that the expansion of the universe is decelerated in the past and is accelerated at present with value
of $q_0=-0.582\pm 0.059$. The decelerated to accelerated phase transition happen at $z\sim 0.76$.
Finally, Figure (\ref{Model D}) shows the results for jerk parameter $j(z)$ obtained from our kinematic analysis,
where at later times we can appreciate a deviation from $\Lambda$CDM (dashed black curve). Although the Fluid/Gravity model is
still in agreement with the standard model and the latter cannot be discarded since it is within the region of propagation of error
at one sigma (gray region).

\section{Conclusion}
In this paper, we have used fluid/gravity duality to find a holographic dark energy model.
This dark energy model was dual to a
five dimensional Schwarzschild black hole.
We were able to find the correct behavior  for this holographic model from its dual
description.   We were thus able to relate the parametric space of the holographic dark energy
model to its dual description. We then analyzed the thermodynamical stability  of this
five dimensional Schwarzschild black hole, and use the fluid/gravity   duality to analyze
the behavior of the dark energy model.   Finally, we used
cosmological observational data to constraint the parametric space of this holographic dark energy
model.\\
Indeed we obtained equation of state by using pressure and density of the holographic fluid. Comparing with dark energy equation of state give us a time-dependent radius corresponding to the time-dependent dark energy model.\\
Then, we considered obtained dark energy model and studied thermodynamics to obtain some thermodynamical parameters like Helmholtz free energy, entropy and specific heat. By using sign of the specific heat and sound speed we found that the model may be unstable.\\

In this work, we  analyzed a specific kind of dark energy models.
It would be interesting to analyze such models using some more generalized
equation of state for a dark energy model \cite{31, 32, 33}. We also analyzed the thermodynamic
stability for the dual description of this model. It would be interesting to analyze the effects
of thermal fluctuations of such a dual description, and then analyze the duality using
such some corrected thermodynamics. It may be noted that thermal fluctuations can change the behavior
of thermodynamical systems
\cite{34,35,36,37,38,39,40,41,42,43,44,45,46,47}, and so it is expected to have direct
effect on the stability of this system.

It would also be interesting to analyze this holographic  formalism for
other  models of dark energy, such as the  generalized Chaplygin gas
\cite{48,49,50,51,52,No3},  generalized cosmic Chaplygin gas \cite{53, 54, 55},
modified Chaplygin gas \cite{56,57,58,59,60,61,61-1},  modified cosmic Chaplygin gas
\cite{62,63,64,65},  extended Chaplygin gas \cite{66,67,68,69,70}. It is interesting to note that it is possible to use the
extended Chaplygin gas models   equation of state \cite{27, 28, 29, 30},
\begin{equation}\label{39}
p=-\frac{B}{\rho^{\alpha}}+A_{1}\rho+A_{2}\rho^{2}+\cdots.
\end{equation}
and analyze the stability of such systems. It would be interesting to use such
equations of states and analyze the  stability of
dark energy model using  a dual description. It is expected that
certain problems with certain values of parameters in this theory can be resolved by
using such extended theories. It would also be interesting to analyze the holographic
dual to such dark energy models, and constraint such models from observations.

\section{Acknowledgments}
The authors would like to thank Rafael C. Nunes for useful comments. The authors thank CNPq (Conselho Nacional de Desenvolvimento Cientifico
e Tecnologico), Brazilian scientific support federal agency, for partial financial support, Grants numbers 302155/2015-5, 302156/2015-1
and 442369/2014-0. E.M.C.A. thanks the hospitality of Theoretical Physics Department at Federal University of Rio de Janeiro (UFRJ),
where part of this work was carried out.

\end{document}